%% file: nucl8.tex
\documentstyle[prl,aps,twocolumn,epsfig]{revtex}
\input{title2.tex}

\newcommand{\dd}{{\rm d}}
\newcommand{\ee}{{\rm e}}
\newcommand{\eqref}[1]{(\ref{#1})}
\newcommand\mean[1]{{\big<#1\big>}}

\newcommand{\tl}{{\bf t}_{\mu}}

\newcommand{\tz}{{\bf t}_{0}}
\newcommand{\pr}[1]{{\cal P}[#1]}

\begin{document}
\draft 
\title{Dynamics of Kinks: Nucleation, Diffusion and Annihilation} 

\author{Salman Habib$^{\dagger}$ and Grant Lythe$^{\star}$}
\preprint{LA-UR 99-1618} 
\address{$^{\dagger}$T-8, Theoretical Division, MS B285, Los Alamos
National Laboratory, Los Alamos, New Mexico 87545}
\address{$^{\star}$Center for Nonlinear Studies, Los Alamos National
Laboratory, Los Alamos, New Mexico 87545}

\date{\today} 

\abstract{\small{ We investigate the nucleation, annihilation, and
dynamics of kinks in a classical $(1+1)$-dimensional $\phi^4$ field
theory at finite temperature. From large scale Langevin simulations,
we establish that the nucleation rate is proportional to the square of
the equilibrium density of kinks. We identify {\em two} annihilation
time scales: one due to kink-antikink pair recombination after
nucleation, the other from non-recombinant annihilation. We introduce a
mesoscopic model of diffusing kinks based on ``paired'' and
``survivor'' kinks/antikinks. Analytical predictions for the dynamical
time scales, as well as the corresponding length scales, are in good
agreement with the simulations.}}

 \maketitle2
\narrowtext

Many extended systems have localized coherent structures that maintain
their identity as they move, interact and are buffeted by local
fluctuations. The statistical mechanics of
these objects has diverse applications, {\em e.g.}, in condensed matter
physics \cite{SS}, biology \cite{PB}, and particle physics
\cite{KRS}. The model to be studied here is a kink-bearing $\phi^4$
field theory in $(1+1)$ dimensions, popular because its properties are
representative of those found in many applications. Static equilibrium
quantities of this theory, such as the kink density and spatial
correlation functions, are now well understood and recent work has
shown that theory and simulations are in good agreement
\cite{AHK,HKS,bhl}. However, dynamical processes, both close to and
far out of equilibrium, are much less well understood. Questions
include: What is the nucleation rate of kink-antikink pairs? How is an
equilibrium population maintained?  How do these dynamical processes
depend on the temperature and damping? These questions, among others,
are the subject of this Letter.

We introduce and analyze below a simple model of kink diffusion and
annihilation that predicts the nucleation rate and provides a picture
of the physical situation, including the existence of multiple time
and length scales. We also carry out high resolution numerical
simulations.  As one consequence of our work, we are able to settle a
recent controversy as to whether the nucleation rate of kinks in an
overdamped system is proportional to $\exp(-2E_k\beta)$ \cite{BC} or
$\exp(-3E_k\beta)$ \cite{HMS} in favor of the first result ($E_k$ is
the kink energy and $\beta=1/k_BT$).

We consider the dynamics of the $\phi^4$ field obeying the following
dimensionless Langevin equation \cite{AHK}:
\begin{equation}
\partial^2_{tt}\phi=\partial^2_{xx}\phi+\phi(1-\phi^2) 
-\eta\partial_t\phi + \xi(x,t),
\label{spde}
\end{equation}
with the fluctuation-dissipation relation enforced by
\(\mean{\xi(x,t)\xi(x',t')}= 2\eta\beta^{-1}\delta (x-x')\delta
(t-t')\). We perform simulations on lattices typically of \(10^6\)
sites, using a finite difference algorithm that has second-order
convergence to the continuum \cite{bhl}. Typical values of the grid
spacing and time step are \(\Delta x = 0.4\) and \( \Delta t = 0.01\).

\begin{figure}
\centerline{\epsfig{figure=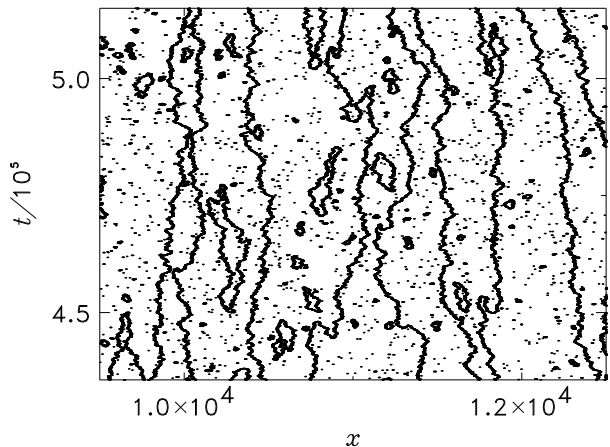,height=6.5cm,width=8.3cm,angle=0}} 
\caption{Timelines of kinks and antikinks: A small spacetime portion of one
 numerical solution is shown. Many recombinant nucleation-annihilation
events are barely visible, forming small closed
loops. \protect\(\beta=7\), \(\eta=1\).}
\label{spacetime}
\end{figure}

At zero temperature, the static kink solution centered at $x=x_0$ is
\(\phi_k(x) = k(x-x_0)\) where \(k(x)=\tanh(x/\sqrt{2})\); the
corresponding antikink solution is \(\phi_a(x) = -k(x-x_0)\). Because
there are only two potential minima, kinks alternate with antikinks on
the spatial lattice. Imposing periodic boundary conditions constrains
the number of kinks and antikinks to be equal. During the time
evolution, we identify kinks and antikinks individually and follow the
``lifeline'' of each kink or antikink (Fig. \ref{spacetime}).

Equilibrium properties of one-dimensional systems, such as the free
energy density and the correlation function \(\mean{\phi(0)\phi(x)}\),
can be calculated using the transfer integral method \cite{ssf}. The
calculation is exact, although one typically must evaluate eigenvalues
of the resulting Schr\"odinger equation numerically. When the on-site
potential has the double-well form, as is the case here, one part of
the free energy density at low temperature can be interpreted as due
to kinks, forming a dilute gas with density \cite{ssf}:
$\rho_k\propto\sqrt{E_k\beta}\exp(-E_k\beta)$. This WKB approximation
is consistent with recent simulations at $\beta > 6$, where
unambiguous identification of kinks is possible \cite{AHK}.

\begin{figure}
\centerline{\epsfig{figure=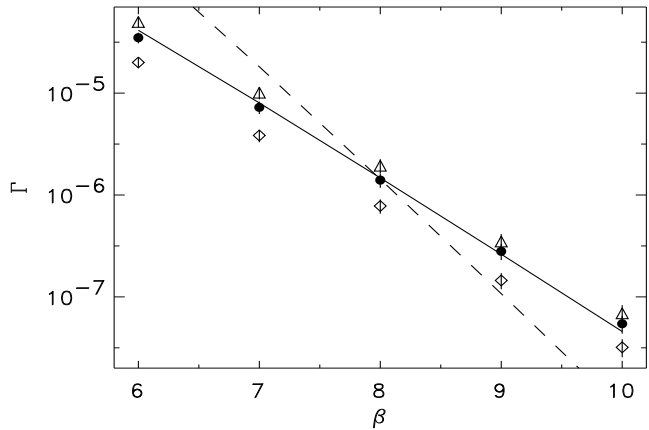,height=6.1cm,width=8.5cm,angle=0}}  
\caption{Nucleation rate, measured from numerical solution of
\eqref{spde} with tracking of timelines. The rate, $\Gamma$, of
production per unit length per unit time of kink-antikink pairs is
shown versus $\beta$ for three values of $\eta$: $\eta=0.2$
(triangles), $\eta=1$ (filled circles) and $\eta=5$ (diamonds). The
solid line is $\Gamma = \rho_k^2$  and the dashed line is the best fit
to  $\Gamma = a\rho_k^3$ for \(\eta=1\).
}
\label{nucl}
\end{figure}

An equilibrium density of kinks is maintained by a dynamical balance
of nucleation and annihilation of kink-antikink pairs (Fig.
\ref{spacetime}).  The dependence of the nucleation rate \(\Gamma\) on
temperature and damping, however, is not directly calculable from the
transfer integral; nor are unambiguous results for symmetric
potentials available from saddle-point calculations \cite{land}. While
analogy with the Kramers' problem suggests \(\Gamma \propto
\exp(-2\beta E_k)\) \cite{kram}, the relationship \(\Gamma \propto
\exp(-3\beta E_k)\) has also been suggested \cite{HMS}. Our direct
counting of nucleation events establishes that their rate is
proportional to the square of the equilibrium density, that is
\(\Gamma \propto \exp(-2\beta E_k)\) (Fig. 2). Below we show how this
relation can be understood from a mesoscopic model of diffusing kinks
with paired nucleation.

At equilibrium, the nucleation rate is related to the mean kink
lifetime \(\tau\) by \(\rho_k = \Gamma \tau\).  Previous attempts to
evaluate \(\Gamma\) numerically \cite{numprev,BD} have proceeded by
counting the number of kinks \(n_k(t)\) and assuming exponential decay
of $\langle n_k(t+\tau)n_k(t)\rangle$. Unfortunately this approach
provides no information on the underlying processes, and yields
incorrect results if kinks are not properly identified on the
lattice. In particular, results that appeared to support
$\Gamma\propto \exp(-3\beta E_k)$ were performed at temperatures too
high for accurate computation of $\langle n_k(t+\tau)n_k(t)\rangle$
\cite{numprev}.

\begin{figure}
\centerline{\epsfig{figure=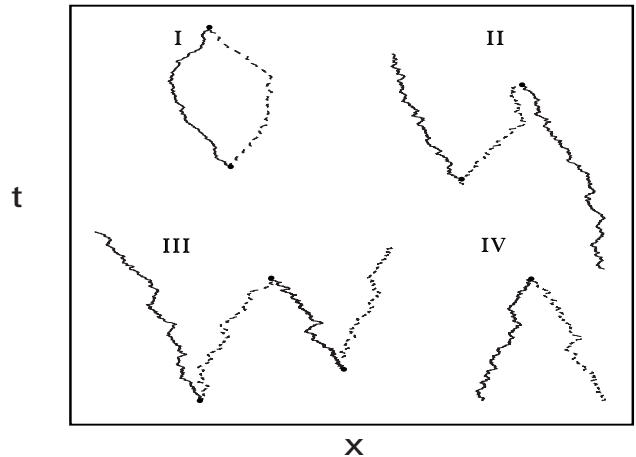,height=6.1cm,width=8.3cm,angle=0}} 
\vspace{.2cm}
\caption{
The four kink-antikink annihilation processes.
 I: Recombination of a paired kink and antikink; II:
Annihilation of a survivor with a paired kink/antikink; III
Annihilation of a kink and an antikink from two neighboring nucleation
sites (producing a survivor kink and antikink); IV Annihilation of two
survivors.}
\label{diags}
\end{figure}

Because we identify individual nucleation events and follow
individual kink lifelines, we can distinguish ``paired'' kinks (whose
partner antikink is still alive) from ``survivor'' kinks (whose
partner has been killed). We also distinguish and measure the
contributions to the annihilation rate from the recombinant and
various non-recombinant mechanisms (Fig. \ref{diags}). The most
frequent annihilation event is recombination of a recently-nucleated
pair (designated I in Fig. \ref{diags}) \cite{BC}. However, the
``survivor'' kinks that remain after a non-recombinant annihilation
event (II or III) have a longer mean lifetime.

At finite temperature, the mean-squared displacement of an isolated
kink is given by \(\mean{{\bf X}_t^2}=2Dt\).  The diffusivity $D$ can
be estimated by using the zero-temperature kink as an ansatz in the
equation of motion~\eqref{spde}, yielding
\(D\simeq(E_k\beta\eta)^{-1}\) \cite{kinkd}, where $E_k=\sqrt{8/9}$
for a static kink.  Corrections to \(D\), arising because of
fluctuations in the kink shape, are proportional to \(\beta^{-2}\) and
subdominant in the temperature range considered here.

Our numerical observations, in particular that kink-antikink
collisions at moderate to large damping always result in annihilation,
motivate us to introduce the following mesoscopic model of kink
dynamics: (i) kink-antikink pairs are nucleated at random times and
positions with initial separation \(b \ll \rho_k^{-1}\); (ii) once
born, kinks and antikinks diffuse independently with diffusivity
\(D\); (iii) kinks and antikinks annihilate on collision. The
separation between a kink and its partner performs Brownian motion
with diffusivity \(2D\).  Thus, if only recombinate annihilation (I in
Fig. \ref{diags}) were allowed, the time \(\tz\) between nucleation
and annihilation would have the density \( \frac{\dd}{\dd t}\pr{\tz <
t} =bt^{-\frac32}(8\pi D)^{-\frac12}\exp(-\frac{b^2}{8Dt}) \)
\cite{kands}.

To analyse our model, we use the following approximation for
non-recombinant annihilation: as long as both members of a pair are
alive, there is a constant probability \(\mu\) per unit time of a
member being struck and ``killed'' by an outsider, {\em i.e.} of an event
II or III. Thus, to each pair we assign a killing time
\(\tl\), distributed according to \(\pr{\tl>t}=\exp(-\mu t)\).
Non-recombinant annihilation happens with probability \(\pr{\tl<\tz} =
1-\exp(-b\nu)\), where \(\nu^2=\frac{\mu}{2D}\) \cite{dandj,hllm}.
The killing rate \(\mu\) depends on the density of kinks; we estimate
it as follows. A new-born pair finds itself in a domain between an
existing kink and antikink of typical length $1/(2\rho_k)$. The mean
time for a diffusing particle to exit the region is proportional to
\((2D\rho_k^2)^{-1}\). Therefore, let
\begin{equation}
   \mu = 2D\alpha^2\rho_k^2.  
\label{mu} 
\end{equation} 
The value of the dimensionless factor was obtained from 
numerical measurements of length and timescales (see Figures 
\ref{bubt} and \ref{fx} below): we estimate \(\alpha \simeq 8\).

Let \(R(t) = \frac{\dd}{\dd
t}\pr{\tz<t|\tz<\tl}\). Then
\begin{equation}
R(t) = N(b)\exp(-\frac{b^2}{8Dt})t^{-\frac32}\exp(-\mu t),
\label{gemlife}
\end{equation}
where \(N(b) = b(8\pi D)^{-\frac12}\exp(\nu b)\).  In Fig.
\ref{bubt} we plot \eqref{gemlife} and a histogram of values of
\(\tz\) obtained from a large numerical solution of 
(\ref{spde}). The behavior \(R(t)\propto t^{-\frac32}\) is
characteristic of Brownian excursions \cite{kands}.

\begin{figure}
\centerline{\epsfig{figure=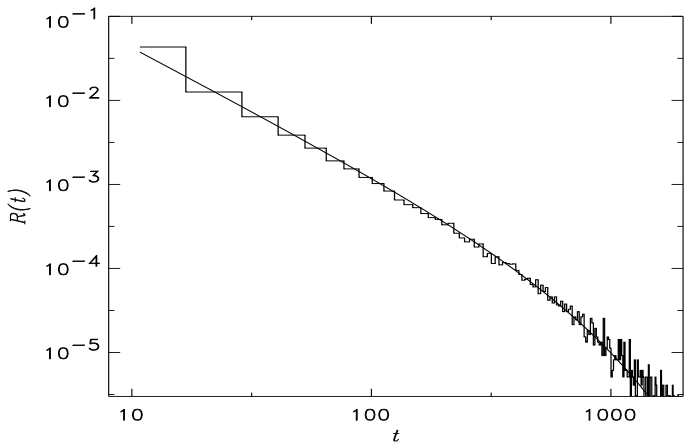,height=6.1cm,width=9cm,angle=0}} 
\caption{Recombination time density. Only recombinant histories (I in
Fig. \ref{diags}) are counted. The solid line is
\protect\eqref{gemlife}. The histogram is from a simulation of
\protect\eqref{spde}. \protect\(\beta=6\), \(\eta=1\).}
\label{bubt}
\end{figure}

Our mesoscopic model has two timescales \cite{hllm}:
\begin{eqnarray}
\tau_0 &=& \mean{\tz|\tz<\tl}=\frac{b}{2\sqrt{\mu D}}~, 
\label{meant0}\\
\tau_{\mu} &=& \mean{\tl|\tl<\tz}
=\frac1{\mu}\left(1-\frac{\nu b}{2}\frac{1}{\ee^{\nu b}-1}
\right)~.
\label{meantmu}
\end{eqnarray}
The mean recombination time \eqref{meant0} depends on \(b\); in
contrast \(\tau_{\mu}\) has a non-zero limit for \(\nu b\to 0\):
$\tau_{\mu}\to 1/(2\mu)$.  With the approximation that a  
``survivor'' kink/antikink has the same probability per unit time,
\(\mu\), of collision and death, the {\em mean lifetime of a kink or
antikink} is given by
\begin{equation}
\tau=\tau_0\ee^{-\nu b}+\tau_{\mu}(1-\ee^{-\nu b})
+\frac1{2\mu}(1-\ee^{-\nu b}).
\label{prev}
\end{equation}
As \(\rho b\to 0\), \( \tau\to (3/2){b}(2\mu D)^{-1/2} 
= (3/4){b}(\alpha D \rho)^{-1} \).
In the same limit, combining \eqref{mu} and \eqref{prev} yields
 the prediction that the nucleation rate is proportional to the 
square of the equilibrium density:
\begin{equation}
\Gamma = \rho_k/\tau = \frac{4}{3b} D \alpha  \rho_k^2,
\label{nucl2}
\end{equation}

The relation \(\Gamma\propto\rho_k^2\) is also found in the
discrete-space Ising model \cite{racz}.
 In contrast, the nucleation rate is proportional to \(\rho_k^3\)
in systems where nucleation does not occur in pairs
\cite{racz,katja}. The latter scaling was incorrectly predicted for
the \(\phi^4\) system \cite{HMS}, from an estimate of the annihilation
rate that does not take into account paired nucleation. (In the
\(\phi^4\) system one does, however, find that the rate of
survivor-survivor annihilation events -- IV in Fig. \ref{diags} --
is proportional to \(\rho_k^3\).)  In the \(\phi^4\) SPDE, the
parameters \(D\), \(\Gamma\) and \(b\) have in general a weak
(non-exponential) temperature dependence.  The lengthscale \(b\) is of
the same order as the width of an isolated kink.

\begin{figure}
\centerline{\epsfig{figure=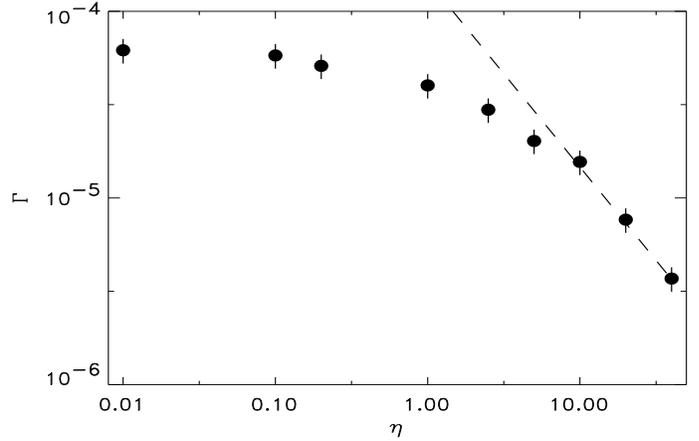,height=6.1cm,width=9cm,angle=0}} 
\caption{ Nucleation rate versus damping for \protect\(\beta=6\). The
dashed line has slope $1/\eta$.}  
\label{nucleta}
\end{figure}

In Fig. \ref{nucleta} we plot the measured nucleation rate versus
the damping coefficient at fixed temperature. The nucleation rate is
proportional to \(\eta^{-1}\) for \(\eta\gg 1\) [in agreement with
(\ref{nucl2})] and appears to plateau for \(\eta\to 0\). At low
damping, however, direct measurement of the nucleation rate is
problematic because kink-antikink collisions
may result in single or multiple bounces rather than
immediate annihilation \cite{csw}.

We now turn to the lengthscales in the system. A
histogram of distances between neighboring kinks and antikinks is
well-approximated by an exponential with characteristic length
\((2\rho_k)^{-1}\).  This simple form results from the
cancellation of the tendency of paired kinks to be closer together than
\((2\rho_k)^{-1}\) with the opposite tendency of survivor kinks.  In
Fig. \ref{fx} we plot \(f(x)=\)(number of occurrences of
separation \(\in (x,x+\dd x)\))\(/(L\,\dd x) \).  We also construct the
 histogram for the separations of {\em only paired} kinks and
 antikinks. The dashed curve is the probability of being at \(x\),
averaged over the lifetime, for a Brownian motion killed at \(x=0\)
and at rate \(\mu\) \cite{dandj}:
\begin{equation} 
   l(x)=N\exp(-\nu x)=N\exp(-\alpha\rho_k x).  
\label{fxsx}
\end{equation}
\begin{figure}
\centerline{\epsfig{figure=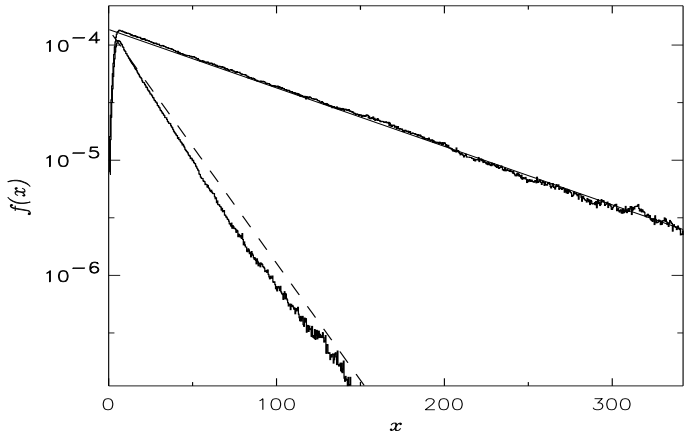,height=6.1cm,width=9cm,angle=0}} 
\caption{Kink-antikink spacing. In the upper part of the plot, a
histogram of all next-neighbor spacings is plotted with
\protect\((2\rho_k)^2\exp(-2\rho_k x)\) (solid line).  The lower part
of the plot is a histogram of kink-antikink spacings only from paired
kinks; the dashed line is \protect\((2\rho_k)^2\exp(-8\rho_k x)\).
\(\beta=6\), \(\eta=1\).}
\label{fx}
\end{figure}

The classification of kinks into paired kinks and survivors, with the
approximation that kinks have a constant probability \eqref{mu} per
unit time of non-recombinant annihilation, allows us to construct a
macroscopic rate theory for the two densities \(n_p(t)\) (paired
kinks) and \(n_s(t)\) (survivor kinks).  The equation for \(n_p(t)\)
has a positive term due to nucleation and a negative term inversely
proportional to the lifetime of pairs, \eqref{meant0}. The terms in
the equation for \(n_s(t)\) correspond to processes III and IV in
Fig. \ref{diags}. Note that process II does not change the number of
survivor kinks. We obtain
\begin{eqnarray}
   \dot n_p =& \Gamma - 2b^{-1}\alpha D (n_s+n_p)\,n_p\cr
   \dot n_s =& D\alpha^2(n_s+n_p)^2(n_p-2n_s).
\label{ratee} 
\end{eqnarray}
The steady state solution of \eqref{ratee} gives the relationship
\eqref{nucl2} between \(\Gamma\) and the equilibrium kink density.
Nonequilibrium dynamics are also correctly described: if \(\Gamma=0\),
the paired density quickly decays and, for late times, \(\dot n_s
\propto n_s^3\), in agreement with an exact result for the survival
probability in the diffusion-limited reaction A\(+\)A\(\to\)0
\cite{tandc}.  While not exact, \eqref{ratee} illustrate that at least
two coupled equations are necessary to capture the two timescales in
the dynamics: no single rate equation can suffice.

We have benefited from discussions with Kalvis Jansons, Eli Ben-Naim and
Vincent Hakim. Computations were performed at the National Energy
Research Scientific Computing Center (NERSC), Lawrence Berkeley
National Laboratory.

\end{document}

%% file: title2.tex
\catcode`\@=11

\def\maketitle2{\par 
\begingroup
\let\cite\@bylinecite
\def\thefootnote{\fnsymbol{footnote}}%
\twocolumn[\@maketitle2\vskip2pc]%
\thispagestyle{plain}\@thanks
\endgroup
\def\thefootnote{\arabic{footnote}}%
\setcounter{footnote}{0}%
\let\maketitle2\relax \let\@maketitle2\relax
\let\@thanks\relax \let\@authoraddress\relax \let\@title\relax
\let\@date\relax \let\thanks\relax \let\@abstract\relax 
\let\@pacs\relax}

\def\abstract#1{\gdef\@abstract{{\par 
\bgroup
\ifdim\prevdepth=-1000pt \prevdepth0pt\fi
\hsize\columnwidth
\dimen0=-\prevdepth \advance\dimen0 by17.5pt \nointerlineskip
\small\vrule width 0pt height\dimen0 \relax}{~~}#1\egroup}}

\def\pacs#1{\gdef\@pacs{{\par 
\bgroup
\hsize\columnwidth \parindent0pt
\ifdim\prevdepth=-1000pt \prevdepth0pt\fi
\dimen0=-\prevdepth \advance\dimen0 by20pt\nointerlineskip
\egroup} PACS numbers:~#1}}

\def\@maketitle2{
\@preprint
\@title
\ifdim\prevdepth=-1000pt \prevdepth0pt\fi
\@authoraddress
\@date
\begin{list}{}{\leftmargin=0.10753\textwidth \rightmargin=\leftmargin
\itemsep=1pc\partopsep=-1pc}
\item\@abstract
\item\@pacs
\end{list}
}

\catcode`\@=12